\documentclass[twocolumn,amsmath,amssymb,floatfix,groupedaddress,nofootinbib]{revtex4-1}

\usepackage{graphicx}
\usepackage{amssymb}
\usepackage{amsmath}
\usepackage{rotating}
\usepackage{hyperref}

\usepackage[usenames,dvipsnames]{xcolor}

\DeclareMathAlphabet\mathbfcal{OMS}{cmsy}{b}{n}
\definecolor{darkgreen}{cmyk}{0.85,0.2,1.00,0.2} 
\definecolor{purple}{cmyk}{0.5,1.0,0,0}

\newcommand{\som}{S_8}

\newcommand{\Oc}{\Omega_c}
\newcommand{\Ob}{\Omega_b}

\newcommand{\LCDM}{$\Lambda$CDM}
\newcommand{\LCDMr}{$r\Lambda$CDM}
\newcommand{\LCDMn}{$\nu\Lambda$CDM} % new LCDM
\newcommand{\LCDMrn}{$\nu r\Lambda$CDM} % newer LCDM
\newcommand{\M}{C}
\newcommand{\EU}{EC}
\newcommand{\LU}{CL}
\newcommand{\ELU}{ECL}
\newcommand{\midd}{central}

\def\barray{\begin{array}} 
\def\earray{\end{array}}
\def\be{\begin{equation}}
\def\ee{\end{equation}}
\def\ben{\begin{equation} \nonumber}
\def\een{\end{equation}}
\def\ban{\begin{eqnarray*}}
\def\ean{\end{eqnarray*}}
\def\ba{\begin{eqnarray}}
\def\ea{\end{eqnarray}}

\def\neff{$N_{\rm eff}$}
\def\dneff{$\Delta N_{\rm eff}$}

\def\({\left(}
\def\){\right)}

\begin{document}

\title{Neutrinos help reconcile Planck measurements with both Early and Local Universe}
%\date{}                                           % Activate to display a given date or no date
\author{Cora Dvorkin}
\email{cdvorkin@ias.edu}
\affiliation{Institute for Advanced Study, School of Natural Sciences, Einstein Drive, Princeton, NJ 08540, USA}
\author{Mark Wyman}
\affiliation{New York University, Center for Cosmology and Particle Physics, New York, NY 10003}
%\email{mark.wyman@nyu.edu}
\author{Douglas H. Rudd}
\affiliation{Kavli Institute for Cosmological Physics, Research Computing Center, University of Chicago, Chicago, Illinois 60637, U.S.A}
%\email{drudd@uchicago.edu}
\author{Wayne Hu }
%\email{whu@background.uchicago.edu}
\affiliation{Kavli Institute for Cosmological Physics,  Department of Astronomy \& Astrophysics,  Enrico Fermi Institute, University of Chicago, Chicago, Illinois 60637, U.S.A}

\begin{abstract}
In light of the recent
BICEP2 B-mode polarization detection, which implies a large inflationary tensor-to-scalar ratio $r_{0.05}=0.2^{+0.07}_{-0.05}$,
we re-examine the evidence for an extra sterile massive neutrino, originally invoked to account for the tension between the cosmic microwave background (CMB) temperature power spectrum and  local measurements of the expansion rate $H_0$ and cosmological structure.   With only the standard active neutrinos and power-law scalar spectra, this detection is in tension with the upper limit of 
$r<0.11$ (95\% confidence) from the lack of a corresponding low multipole excess in the temperature
anisotropy from gravitational waves.   An extra sterile species with the same energy density as is needed to reconcile the CMB data with $H_0$  measurements can also alleviate this new tension.
By combining data from the Planck  and ACT/SPT temperature spectra,
WMAP9 polarization, $H_0$, baryon acoustic oscillation and local cluster abundance
measurements  with  BICEP2 data, we find 
the joint evidence for a sterile massive neutrino increases to \dneff$=0.98\pm 0.26$ for the effective number and
$m_s= 0.52\pm 0.13$ eV  for the effective mass or $3.8\sigma$ and $4\sigma$ evidence respectively. 
We caution the reader that these results correspond to a joint statistical evidence and, in addition, astrophysical systematic errors in the clusters and $H_0$ measurements, and small-scale CMB data could weaken our conclusions.
\end{abstract}
\maketitle

The recent detection of degree scale B-mode polarization in the Cosmic Microwave Background (CMB)  by the BICEP2 experiment \cite{Ade:2014xna}
implies that the inflationary ratio of tensor-to-scalar fluctuations is $r_{0.05}=0.2^{+0.07}_{-0.05}$, a number in significant tension with the upper limit of
 $r<0.11$ at 95\% confidence level from the temperature anisotropy spectrum in the simplest inflationary $\Lambda$CDM cosmology \cite{Ade:2013zuv}.   This conflict occurs because the large angle
 temperature excess implied by the gravitational waves is not observed, and the mismatch 
 cannot be compensated by parameter changes in this highly restricted, seven parameter model.  
 
Aside from the possibility of 
 large astrophysical \cite{Ade:2014xna,Flauger:2014qra,Mortonson:2014bja} or cosmological
\cite{Dent:2014rga,Moss:2014cra,Lizarraga:2014eaa,Mortonson:2010mi,Bonvin:2014xia} foreground and systematic contamination, possible solutions include
extending the inflationary side or the $\Lambda$CDM side of this model.
Inflationary modifications include a large running of the
scalar tilt   \cite{Ade:2014xna}; more explicit features in
the inflationary scalar spectra \cite{Contaldi:2014zua,Miranda:2014wga,Abazajian:2014tqa}; or anticorrelated isocurvature perturbations \cite{Kawasaki:2014lqa}.
In the present work, we instead consider
extensions to the matter content of the $\Lambda$CDM model that can alleviate this early Universe tension.
 
In the $\Lambda$CDM cosmology, recall that the Planck CMB temperature anisotropy
spectrum is also in conflict with measurements
of the local Universe \cite{Wyman:2013lza,Hamann:2013iba,Battye:2013xqa}. 
The $\Lambda$CDM values for the current expansion rate -- or Hubble constant, $H_0$ -- and the abundance of galaxy clusters are
individually in $2-3\sigma$ tension with direct measurements.  While similar tensions existed
in previous CMB data sets (e.g.~\cite{Hou:2012xq}), the Planck results suggest 
a shift in the sound horizon that brings close agreement between CMB-based $H_0$ inferences
and baryon acoustic oscillation (BAO) measurements while disfavoring other possible explanations
related to cosmic acceleration physics.

In this work, we show that both the early Universe and the local Universe
tension with the Planck data may be pointing to the same extension of the $\Lambda$CDM cosmology:
an extra massive sterile neutrino.
Such a neutrino would modify the sound horizon at recombination, which is used to infer distances with both the CMB and BAO,
removing the tension with local measurements of  the Hubble constant.   By changing the relationship between the CMB sound horizon and the
 damping scale, it also leads to an increase in the scalar tilt that suppresses
large angle anisotropy relative to the simpler model.   Finally, if the neutrino carries a mass in the eV range it 
suppresses the growth of structure and hence reduces the number of galaxy clusters
predicted in the local Universe.

In \S \ref{sec:modelsdata} we define the $\Lambda$CDM model along with its tensor and neutrino extensions and group the data
sets by the early and local Universe tensions they expose.   We present results in \S \ref{sec:results} and discuss them in
\S \ref{sec:discussion}.

\section{Models and Data}
\label{sec:modelsdata}

The simplest inflationary $\Lambda$CDM model is characterized by 6 parameters,
$\{\Oc h^2, \, \Ob h^2, \, \tau, \theta_{\rm MC}, A_{s}, n_s \}$. 
Here, $\Oc h^2$ represents the physical cold dark matter (CDM) density, $\Ob h^2$  is the baryon density, $\tau$ is the Thomson optical depth to reionization, $\theta_{\rm MC}$ is a proxy for the angular acoustic scale at recombination, $A_s$ is the amplitude of the initial curvature power spectrum at $k = 0.05$ Mpc$^{-1}$, and $n_s$ its spectral index.  

To these parameters we add  $r$, the tensor-scalar ratio evaluated at $k=0.002$ Mpc$^{-1}$,
and take the tensor tilt to follow the consistency relation $n_t = -r/8$.  
For the neutrino extension we add two new parameters.
The first is the effective number of relativistic species, $N_{\rm eff}$, defined in terms of the relativistic energy density at high redshift
\be
\rho_{\rm r} = \rho_\gamma + \rho_\nu = \left[ 1 + \frac{7}{8} \(\frac{4}{11}\)^{4/3}  \hspace{-6pt} N_{\rm eff} \right] \rho_\gamma.
\ee
In the minimal model, $N_{\rm eff}=3.046$ and so $\Delta N_{\rm eff}=N_{\rm eff}-3.046>0$
indicates the presence of extra relativistic particle species in the early Universe.   We assume that the
active neutrinos have  $\sum m_\nu=0.06$ eV (as suggested by a normal hierarchy and solar and atmospheric
 oscillation measurements \cite{GonzalezGarcia:2012sz}) and any additional contributions 
are carried by a mostly sterile state, with effective mass $m_s$ for
a total neutrino contribution to the energy density today of
\be
\label{omegan}
({94.1 \, \rm eV}) \Omega_\nu h^2 = {(3.046/3)^{3/4} \sum m_\nu+m_s}.
\ee
Thus  $m_s$ characterizes extra non-CDM energy density rather than the true
(Lagrangian) mass of  a neutrino-like particle.  In particular even in the \dneff$\rightarrow 0$ limit,
at fixed $m_s$, the presence of an extra sterile species still
adds an extra energy density component at recombination and still changes inferences based on the sound horizon.   

In the most recent version of CosmoMC \cite{Lewis:2002ah}, a prior on the physical mass of a thermally produced
sterile neutrino is imposed with a value of $m_s/(\Delta N_{\rm eff})^{3/4} < 10{\rm eV}$ to close off a degeneracy between very massive neutrinos and cold dark matter.

We call these extensions
 the 
 \LCDMn\ and \LCDMrn\ models, where the names follow from the additional parameters
introduced by tensors and neutrinos.

\begin{table}[t] \centering
\begin{tabular}{ c | c  }
 & Models\\
\hline
\LCDM & $\{\Oc h^2, \, \Ob h^2, \, \tau, \theta_{\rm MC}, A_{s}, n_s \}$ \\
\LCDMr & \LCDM + r \\
\LCDMn & \LCDM +\neff$+m_s$ \\
\LCDMrn & \LCDM  +\neff$+m_s+r$\\
\multicolumn{2}{c}{} \\
  & Data sets\\
\hline
\M &  \{Planck, WP, SPT/ACT\} \\
\EU & \M +BICEP2 \\
\LU & \M +$H_0$+BAO+Clusters\\\
\ELU &\M +BICEP2 +$H_0$+BAO+Clusters\\
\end{tabular}
\caption {Model and data set combinations. All models include \LCDM\ parameters.
 All data sets include the \midd\ Universe (\M) set.
}  
 \label{tab:datamodel} 
\end{table}

For the data sets, we consider combinations that best expose the separate early and
local Universe tensions with what we refer to as the \midd\ (\M) Universe 
-- the time intermediate between inflation at the late Universe, 
characterized chiefly by the CMB temperature power spectrum.   This \M\ data set is composed of  the Planck 
temperature \cite{Ade:2013zuv}, WMAP9 polarization \cite{Bennett:2012fp}, and ACT/SPT 
 \cite{Das:2013zf,Reichardt:2011yv,Keisler:2011aw} high multipole power spectra.
  For the Planck data analysis we marginalize over 
 the standard Planck foreground parameters \cite{Ade:2013zuv}.
  For the early Universe tension, we 
add to  these what we call the early (E) data set, the BICEP2 BB and EE polarization bandpowers \cite{Ade:2014xna}.
We call the combination of these two the  early-\midd\  (\EU)  Universe data set.

For the local Universe tension, we define the following collection as the local (L) data sets: the  $H_0$ inference from the maser-cepheid-supernovae  distance ladder, $h = 0.738 \pm 0.024$ \cite{Riess:2011yx},  
BAO measurements   
\cite{Anderson:2012sa, Padmanabhan:2012hf, Blake:2011en}, and the X-ray derived cluster abundance using the likelihood code\footnote{This prescription employs the total matter power spectrum.  More recent studies of the cluster abundance indicate that this may somewhat underestimate the neutrino mass at least at lower masses than considered here \cite{Villaescusa-Navarro:2013pva} where the
abundance follows the cold dark matter power spectrum more closely. We continue to adopt this
approach to be conservative with respect to new neutrino physics and provide a continuous limit with CDM at high $m_s$.} of Ref.~\cite{Burenin:2012uy} which roughly equates
to a constraint on $S_8=\sigma_8 (\Omega_m/0.25)^{0.47}=0.813 \pm 0.013$. Note that the BAO data are added here not because they are in tension with Planck (they are not) but because they exclude resolutions of the $H_0$ tension involving exotic dark energy or curvature.
To these we again add the Planck temperature, WMAP9 polarization and the ACT/SPT data sets. 
We call this the \midd-local (\LU)  Universe  data set. Finally, we call the union of this with the \EU\  data the \ELU\ data set.
These model and data choices are summarized in Tab.~\ref{tab:datamodel}.

We analyze these data and models using the Markov Chain Monte Carlo technique 
and the CosmoMC code \cite{Lewis:2002ah}. Our local analysis will be in many ways similar to that performed in Ref.~\cite{Wyman:2013lza} to which we refer the reader for details and robustness checks.

\begin{table*}[t] \centering
\begin{tabular}{ c | c | c| c  }
&  \LCDMrn--\EU &  \LCDMn--\LU & \LCDMrn--\ELU \\
\hline
\dneff\  	&$1.06\pm0.37$ & $0.62\pm0.28$ & $0.98\pm0.26$ \\
$m_s$ [eV]	&$<0.22$ & $0.48\pm0.15$ & $0.52\pm0.13$ \\
$r$ 		& $0.19\pm0.05$ & --	& $0.22\pm0.05$ \\
\hline  
$100\Omega_b h^2$&$2.268\pm0.043$ & $2.267\pm0.028$ & $2.276\pm0.027$\\
$\Omega_c h^2$ &$0.132\pm0.005$ & $0.122\pm0.005$ & $0.127\pm0.004$\\
$100\theta_{\rm MC}$&$1.040\pm0.001$ & $1.041\pm0.001$ & $1.040\pm0.001$\\
$\tau$ 		&$0.100\pm 0.015$ & $0.096\pm0.014$ & $0.097\pm0.014$\\
$\ln (10^{10} A_s)$ & $3.136\pm0.033$ & $3.107\pm0.031$ & $3.117\pm0.030$\\
$n_s$		&$0.999\pm0.017$ & $0.985\pm0.012$ & $1.001\pm0.010$\\
\hline
$h$			& $0.74\pm0.04$ & $0.70\pm0.01$ & $0.72\pm0.01$\\
$S_8$		&$0.89\pm0.03$ & $0.81\pm0.01$ & $0.81\pm0.01$\\
\end{tabular}
\caption {Parameter constraints (68\% confidence level) with various model and data assumptions.  Note that the \LCDMn-\LU\ case is in a different, no tensor model, context than the others which affects parameter interpretations.}  
 \label{tab:results} 
\end{table*}

\section{Results}
\label{sec:results}

We begin by discussing the tension introduced by the BICEP2 data in the \EU\ data set in the \LCDMr\ model and its alleviation in the \LCDMrn\ space independently of the \LU\ data.

\begin{figure}[t!]
\includegraphics[width=\columnwidth]{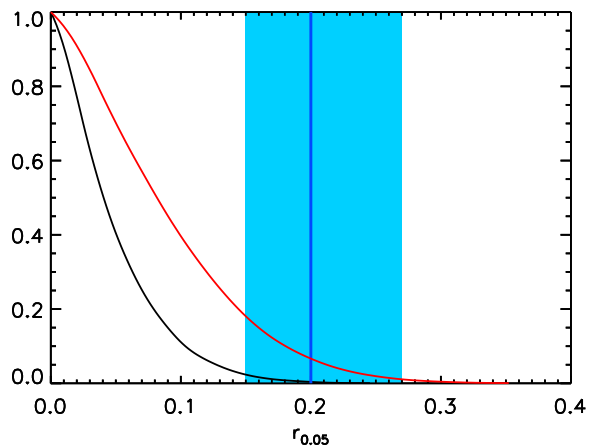}
\caption{\footnotesize BICEP2  measurement of the tensor-scalar ratio $r_{0.05}$ (bands)
compared with the posterior probability distribution of the \M\ data set in the 
\LCDMr\ model space (black curve) and \LCDMrn\ space (red curve).  In the former, the measurement is in strong tension with the posteriors whereas the addition of massive sterile neutrinos in the latter allows high $r$. For both the curves and the band, the tensor-to-scalar ratio is evaluated at a pivot scale of $k=0.05$Mpc$^{-1}$ unlike elsewhere. }
\label{fig:rtension}
\end{figure}

In Fig.~\ref{fig:rtension}, we show the posterior probability distribution of the scalar-tensor ratio
 in these
two models.  In order to compare distributions with the quoted BICEP2 result of 
$r_{0.05}=0.2^{+0.07}_{-0.05}$, we show results for the ratio at $k=0.05\,$Mpc$^{-1}$ unlike
the $0.002\,$Mpc$^{-1}$ value assumed elsewhere.
Note that in \LCDMr\ the \M\ data set imply an upper limit of $r_{0.05}<0.1$ 
consistent with the Planck collaboration analysis \cite{Ade:2013zuv} but in tension with the
E or BICEP2 data.   Moving to the
\LCDMrn\ space, constraints on $r_{0.05}$ weaken and allow $r_{0.05}=0.2$ within the 95\% confidence
limits.

\begin{figure}[t!]
\includegraphics[width=\columnwidth]{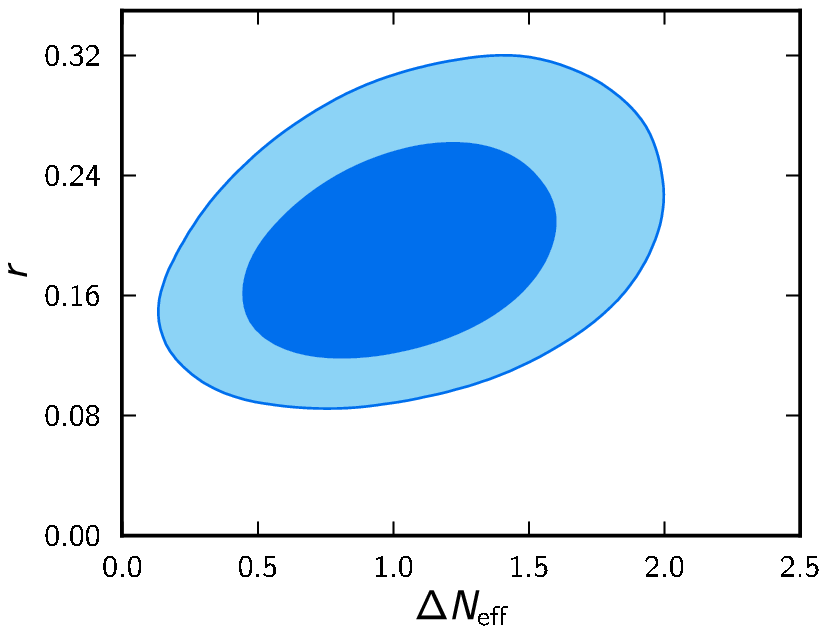}
\caption{\footnotesize Early  Universe tension and neutrinos.   In the \LCDMrn\ parameter space the \EU~data set favors \dneff$>0$ in order to offset the excess large angle temperature anisotropy implied by the high tensor-scalar ratio $r$ (68\%, 95\% contours here and below). This in turn is driven by the degeneracy between \dneff\ and $n_s$ illustrated in Fig. \ref{fig:tilt}. In brief, gravitational waves add power at low $\ell$, requiring larger $n_s$ to compensate. Larger $n_s$ then requires larger \dneff\ to agree with the higher-$\ell$ CMB.}
\label{fig:earlytension}
\end{figure}

\begin{figure}[t!]
\includegraphics[width=\columnwidth]{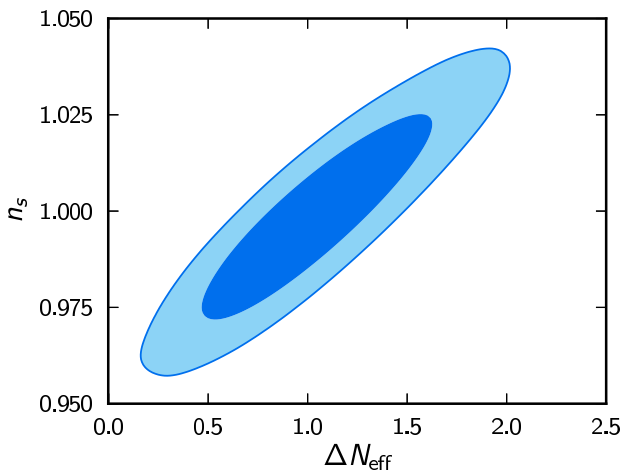}
\caption{\footnotesize    In the \LCDMrn\ parameter space the \EU~data set allows a positive change in the  tilt  when \dneff\ is increased
explaining the mechanism by which the large angle temperature anisotropy is reduced.}
\label{fig:tilt}
\end{figure}

In Fig.~\ref{fig:earlytension} we show the two dimensional $r-$\dneff\ posterior for the \EU\ data and the \LCDMrn\ model.
Note in particular that $r\sim 0.2$ would favor a fully populated \dneff$\sim 1$ extra
neutrino state, while \dneff$=0$ is significantly disfavored (at $2.9\sigma$ once $r$ is marginalized, see Tab.~\ref{tab:results}).    
The origin of this preference is exposed by examining the $n_s-$\dneff\ plane in 
Fig.~\ref{fig:tilt}.   Extra neutrino energy density
at recombination allows a higher tilt and hence removes excess power in the low multipole temperature anisotropy.   For example changing $n_s$ from $0.96$ to $1$
reduces the amount of power at $k=0.002$ Mpc$^{-1}$ relative to 0.05 Mpc$^{-1}$ by
$0.88$, a reduction comparable to the 
amount of temperature power added by tensors when $r=0.2$.

\begin{figure}[t!]
\includegraphics[width=\columnwidth]{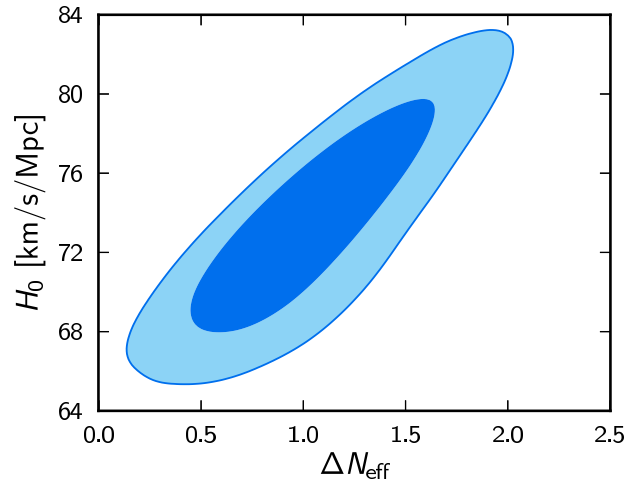}\\
\includegraphics[width=\columnwidth]{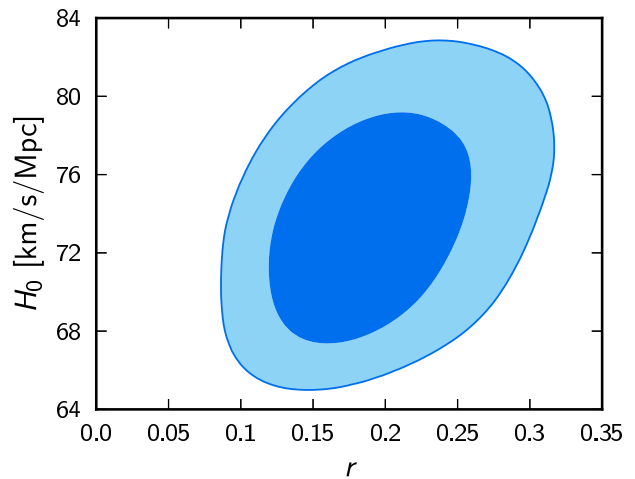} 
\caption{\footnotesize 
By allowing for tensors and neutrinos,  the \EU\ data set in the \LCDMrn\ model
favors higher values for $H_0$.
Note that the actual measurements of $H_0$ are {\it not} imposed here as a prior -- the BICEP2 central value of $r$ in \LCDMrn\ predicts an $H_0$ in concordance with observations.}
\label{fig:EUH0Neutrino}
\end{figure}

\begin{figure}[t!]
\includegraphics[width=\columnwidth]{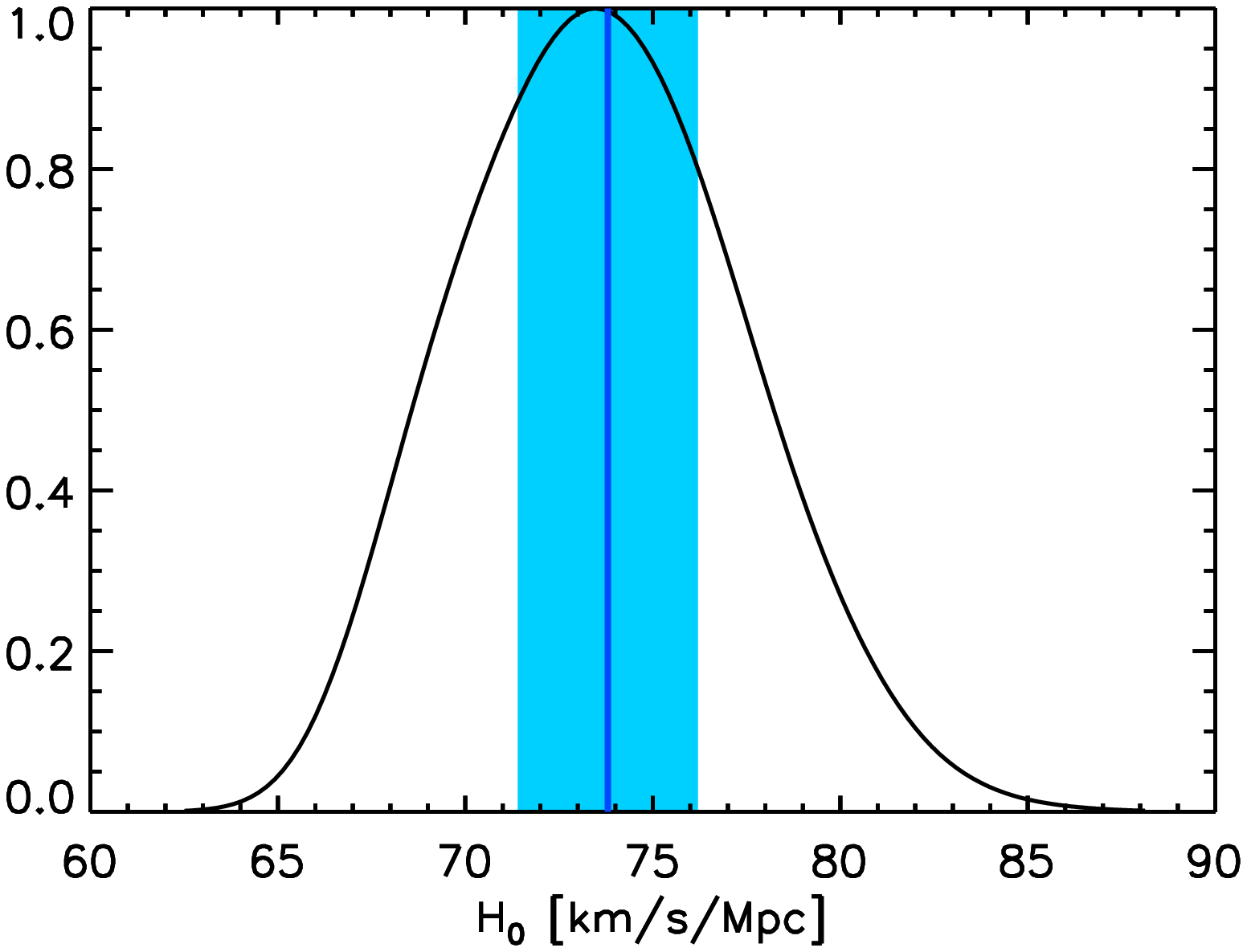}\\
\includegraphics[width=\columnwidth]{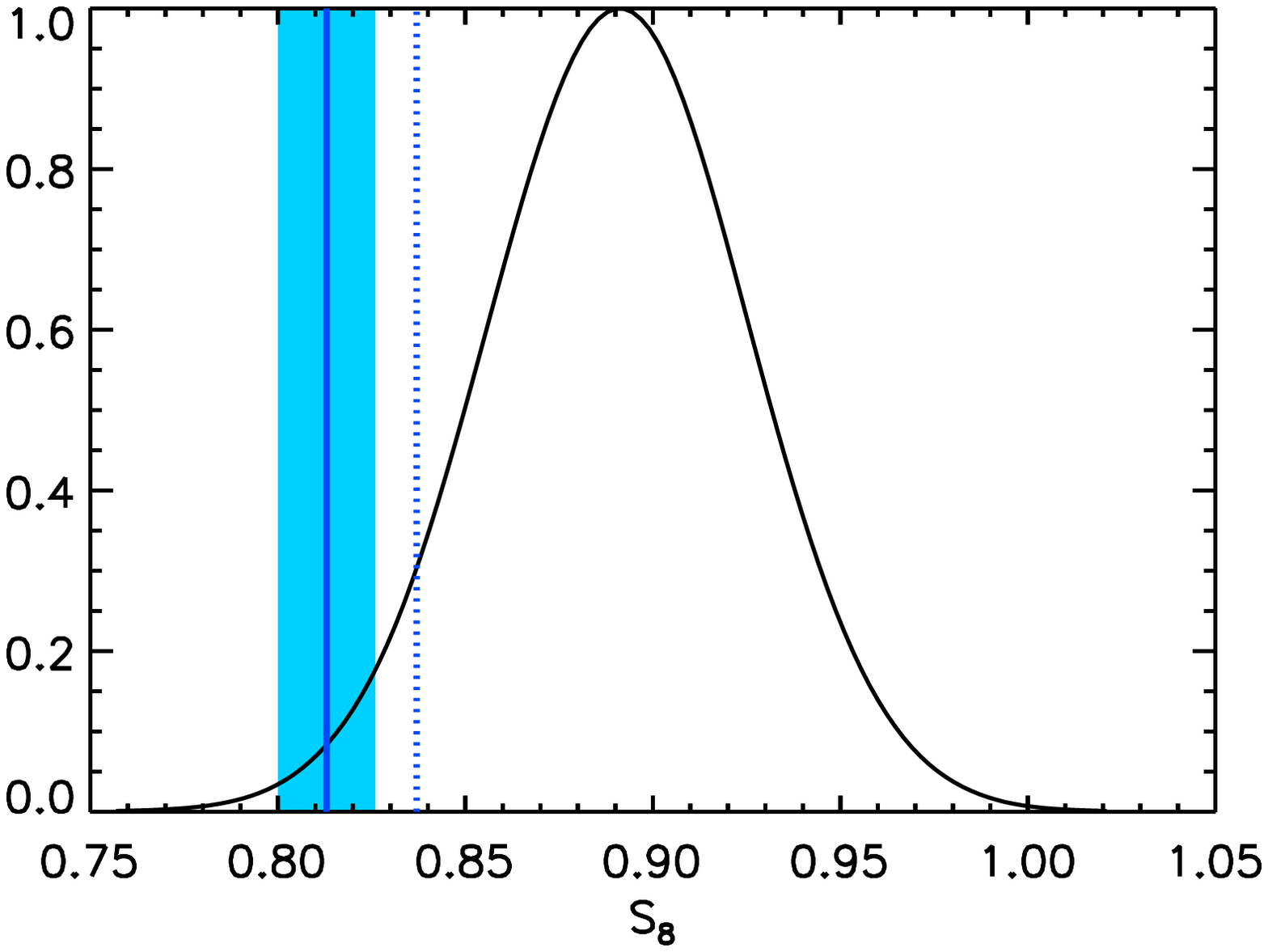} 
\caption{\footnotesize $H_0$ and $S_8$ posterior probability distributions in the
\LCDMrn\ parameter space using early Universe (\EU) data alone compared with the late Universe measurements.  To emphasize, the black 1D posteriors plotted here have been derived without any use of local Universe data. The addition of neutrinos and tensors makes the $H_0$ posterior
fully compatible with measurements (bands) and $S_8$ substantially more compatible, though some residual tension remains.   The dotted line represents the shift in the central value of the cluster
abundance measurements under the assumption of a 9\% systematic increase in cluster masses.}
\label{fig:residualtension}
\end{figure}

This change simultaneously relaxes the CMB-\LCDM\ upper bound on $H_0$,
as can be seen in Fig.~\ref{fig:EUH0Neutrino}.    Extra neutrino energy density at recombination
changes the amount of time sound waves propagate in the CMB-baryon plasma and hence
the standard ruler for CMB and BAO distance measures.

Note that the \EU\ data set does not incorporate late Universe measurements of $H_0$ or
$S_8$.   It is therefore interesting to compare the posterior probability of these parameters
with the actual measurements before combining them into a joint likelihood.    In Fig.~\ref{fig:residualtension}, we show these distributions from the \LCDMrn-\EU\ analysis.    Predictions for $H_0$ in this model context are now fully
compatible with measurements (bands) and correspond to $0.74\pm 0.04$.  The addition of tensors shifts the distribution 
to higher $H_0$ as compared with neutrinos alone (cf.~\cite{Wyman:2013lza}).
The predicted value of the cluster observable is $S_8=0.89\pm 0.03$, so residual tension remains
-- albeit slightly less tension than with neutrinos alone. 
However, this posterior can
accommodate the local observations within the 95\% confidence range.   This tension can be further reduced
if there is a 9\% upward shift in the mass calibration of clusters, which is currently allowed
\cite{Burenin:2012uy}.

\begin{figure}[t!]
\includegraphics[width=\columnwidth]{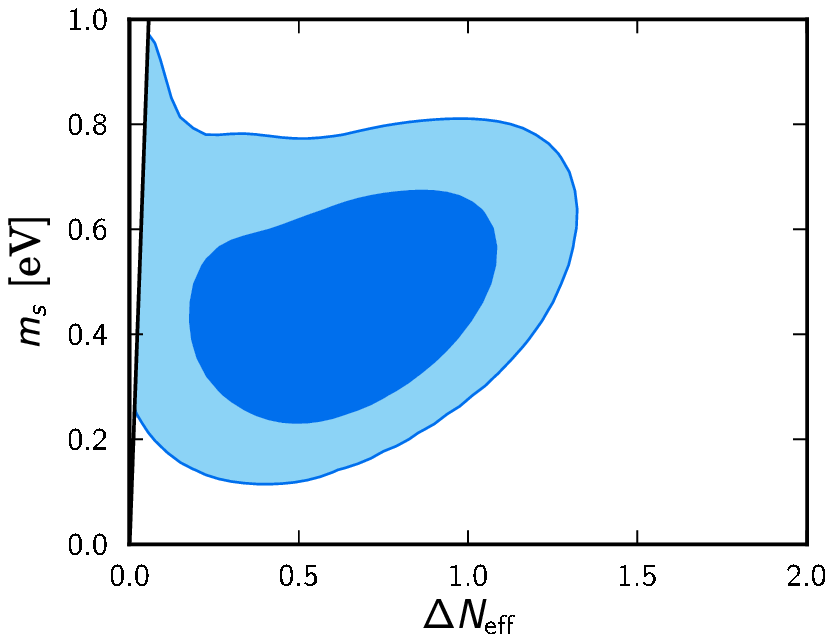}
\caption{\footnotesize Local  Universe tension and neutrinos. The \LU\ data set in the
\LCDMn\ parameter space strongly disfavors the minimal neutrino model with \dneff=0
and $m_s=0$. This agrees with Ref.~\cite{Wyman:2013lza,Hamann:2013iba,Battye:2013xqa}, and comes about because 
local $H_0$ measurements and local cluster abundance measurements add coherently in the direction of preferring new neutrino physics.  Note that the line at  $m_s=10(\Delta N_{\rm eff})^{3/4}\,{\rm eV}$ represents the prior on the physical mass.}
\label{fig:latetension}
\end{figure}

\begin{figure}[t!]
\includegraphics[width=\columnwidth]{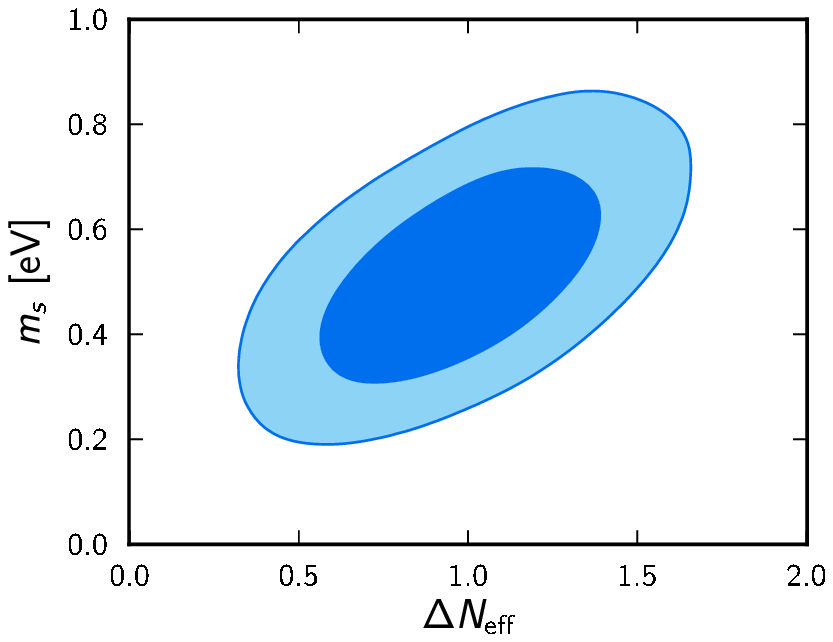}
\caption{\footnotesize Joint evidence for sterile massive neutrino in the combined \ELU\ data set and
\LCDMrn\ model.  The minimal
neutrino model of \dneff$= 0$, $m_s=0$, is rejected at even higher confidence than in the \LU-\LCDMrn\ combination and more strongly favors \dneff$>0$ (cf. Fig. \ref{fig:latetension}). Note also that the same prior on $m_s$ vs. \dneff\ has 
been imposed here as in Fig. \ref{fig:latetension}; we do not shade the excluded region since these contours have
not been distorted by this prior.}
\label{fig:resolution}
\end{figure}

Likewise it is instructive to review
the local tension in the \LU\ data set and its possible resolution in the \LCDMn\ model space
independently of the \EU\ data.   Fig.~\ref{fig:latetension} shows the \dneff$-m_s$ plane.  The \LU\
data imply \dneff$=0.62\pm 0.28$ and $m_s=0.48\pm 0.15$ eV in agreement with  Ref.~\cite{Wyman:2013lza}, strongly excluding the minimal neutrino model at \dneff$=0$ and $m_s=0$.
Without tensors the \M\ component of the \LU\ data set places an upper limit on \dneff\ that disfavors \dneff$=1$.   Thus the Hubble constant inferred is $h=0.70\pm 0.01$, which
is consistent with measurements but on the low side.   Meanwhile,
 the 
cluster abundance data in the \LU\ data set dominates the mass constraint. 
 Note that we do not consider the impact of
systematic errors on the determination of cluster masses here.   In Ref.~\cite{Wyman:2013lza}, it was shown that the preference for a large $m_s$ would only be eliminated
if the masses are underestimated by $\sim 30\%$ which is large compared with the
$9\%$ uncertainty quoted in Ref.~\cite{Burenin:2012uy}.    The full list of parameter
constraints is given in Tab.~\ref{tab:results}.

\begin{figure}[t!]
\includegraphics[width=\columnwidth]{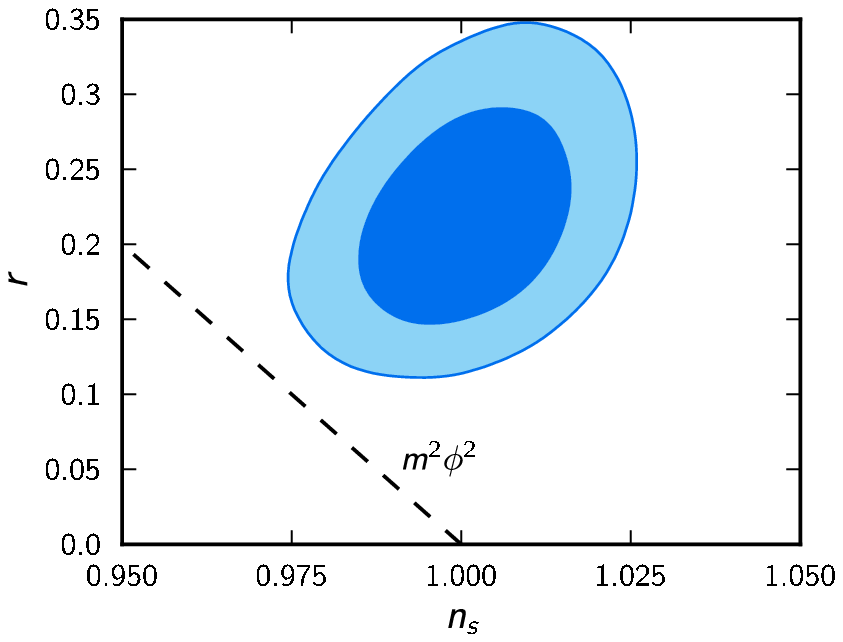}
\caption{\footnotesize Inflationary implications from the 
 combined \ELU\ data set and
\LCDMrn\ model.  Inflationary models with nearly scale invariant tilt and $r\approx 0.2$ 
tensor-to-scalar ratio are favored.   While this region is allowed in slow roll inflation, models
with featureless power law potentials such as $m^2\phi^2$ do not have trajectories (dashed line) that intersect this region.}
\label{fig:inflation}
\end{figure}

\begin{figure}[t!]
\includegraphics[width=\columnwidth]{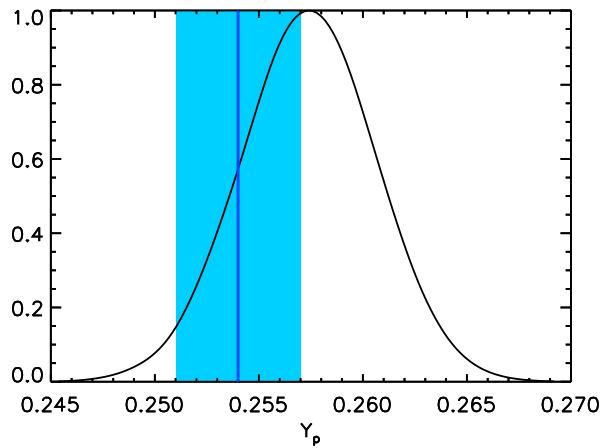}
\caption{\footnotesize Distribution of the helium fraction for the 
 combined \ELU\ data set and
\LCDMrn\ model compared with the BBN measurements (band) \cite{Izotov:2013waa}.}
\label{fig:Yp_distribution}
\end{figure}

Finally we consider the combined early and late Universe (\ELU) data sets  in the full \LCDMrn\ parameter space.  By including tensors in the model, we again both enable a larger
\neff\ and simultaneously fit the tensor results of BICEP2 and $H_0$ measurements.
We find that the preference for an extra massive neutrino has increased in the joint data over the \LU\ case, especially for \dneff.    
As shown in Fig.~\ref{fig:resolution} and Tab.~\ref{tab:results}, \dneff$=0.98\pm 0.26$ and
$m_s= 0.52\pm 0.13$ eV or $3.8\sigma$ and $4\sigma$ evidence respectively.
In particular, adding tensors closes off solutions where $H_0$ is altered by adding
energy density at recombination through $m_s$ at \dneff$\rightarrow 0$ (cf. Figs. \ref{fig:latetension}, \ref{fig:resolution}).  
In addition, the distribution of the helium fraction for this combined data set is consistent with the Big Bang Nucleosynthesis measurements (see Fig. \ref{fig:Yp_distribution}).

These conclusions could weaken if there were systematic problems in the cluster abundance measurements (as suggested, for example, by Ref. \cite{2014MNRAS.438...49R}), $H_0$ measurements \cite{Efstathiou:2013via} or in the small-scale CMB data \cite{Spergel:2013rxa}. Ref. \cite{Leistedt:2014sia} emphasizes that the tension
between  CMB $+$ BAO data and the cluster measurements is not fully resolved
by neutrinos, consistent with Fig.~\ref{fig:residualtension}, and interpret this as 
an indication that the cluster data is not robust.  Even without the cluster data, the neutrino
solution is favored in order to satisfy $H_0$, BAO and BICEP2 data  simultaneously with the 
C data.  In fact it is the agreement with BAO that make the single biggest contribution
to the likelihood improvement from neutrinos (see Tab.~\ref{tab:ML}) and what is limiting the neutrino
explanation is mainly a small tension with SPT/ACT.  Furthermore compared with the $\Lambda$CDM model
that best fits the C data alone, the \LCDMrn\ maximum likelihood model is actually
a better fit to the BAO data itself and improves $H_0$ agreement by
$2\Delta\ln L \approx 6.2$.  The net improvement in maximum likelihood from adding neutrinos with the \ELU\ data sets is $2 \Delta \ln L \approx -24$. 

\begin{table}[htbp!]
\begin{center}
\begin{tabular}{|c|c|c|}
\hline
&  r$\Lambda$CDM (ECL)&  $\nu$r$\Lambda$CDM (ECL)  \\
\hline
Planck  & 18.15&7.37 \\
\hline
WP &-0.34&-1.03 \\
\hline
SPT/ACT&1.03 & 4.85\\
\hline
BAO& 10.56& -0.36\\
\hline
$H_0$&-6.37 & -6.20\\
\hline
\end{tabular}
\end{center}
\caption{Changes in the individual contributions to $2\ln\mathcal{L}$ relative to the $\Lambda$CDM maximum likelihood model with "C" data.}
\label{tab:ML}
\end{table}

We close with a brief mention of the inflationary consequences of our analysis.
In our final analysis, we find $n_s \sim 1$ and $r\sim 0.2$.   
Such models are allowed within the framework of slow roll inflation since they do not have a
large running of the tilt. However, they fall outside the class of simple featureless monomial
potentials (see Fig.~\ref{fig:inflation}).  To achieve nearly exact scale invariance,
an inflationary potential must have its slope and curvature partially cancel in their
effect on the tilt, a less typical situation.  There is a relative 
dearth of such models in the literature (see e.g.~\cite{Martin:2013tda}), although predictions
for such models can be found, for example, in Refs. \cite{Barrow:1993zq,Barrow:2006dh,Czerny:2014qqa}. 

\section{Discussion}
\label{sec:discussion}

We have shown that the tension introduced by the detection of  large amplitude
gravitational wave power by the BICEP2 experiment with temperature
anisotropy measurements is alleviated in a model with an extra sterile neutrino.   
The relativistic energy density required to alleviate this early Universe tension is the same
as that required to resolve the late Universe tension of acoustic distance measures with
local Hubble constant measurements.     Combined they imply a \dneff$=0.98\pm 0.26$.
Note that the ACT/SPT data already limit the upper range of allowed \dneff, and so
this explanation of the early and late Universe tensions can be tested with more data
from high multipole CMB temperature and polarization observations.  Conversely,
it would weaken if extra evidence for alternate inflationary or foreground explanations of the early Universe
tension is found.

By making the sterile neutrino massive, the tension with growth of structure measurements  can simultaneously be alleviated.   The combined constraint on the effective mass is $m_s=0.52\pm 0.13$ eV.
This 
preference for high neutrino mass(es)
is mainly driven by the cluster data set (cf.~Ref.\ \cite{Verde:2013cqa} who find upper limits without clusters).    
Compared with analyses that do not include gravitational waves, the sterile neutrino suggested here has a smaller
expected physical mass, thanks to the generally larger values of \dneff\ that we find 
(recall the thermal conversion formula: $m_s^{\rm th} = m_s/(\Delta N_{\rm eff})^{3/4}$).
At the upper range of \dneff, it is thus somewhat less likely that this sterile neutrino could explain anomalies in short baseline and reactor neutrino experiments 
(see Refs.\ \cite{Conrad:2013mka,Abazajian:2012ys} for reviews).

If future data or analyses lead to increased mass estimates for the
clusters, that change would weaken the preference for a non-zero
$m_s$ by increasing $\som$, giving better concordance with the basic \LCDM\ prediction. However,
the preference can only be eliminated if the systematic shift is
roughly triple the 9\% estimate; that estimate is derived from comparisons of a variety
of X-ray, optical, Sunyaev-Zel'dovich, and lensing observables (see
e.g.~\cite{Rozo:2013hha} for a recent assessment).  

Taken at face value, these results leave us with a potentially very different cosmological standard model. Gravitational waves are 
nearly indisputable evidence for an inflationary epoch, but the lack of a significant primordial tilt compared with $r$ would suggest somewhat unusual inflationary physics
is at play where the impact of the slope and curvature of the inflationary potential partially cancelled
each other.   On the other hand, the new neutrino physics favored here is less
contrived  than that proposed in Ref.~\cite{Wyman:2013lza,Hamann:2013iba,Battye:2013xqa}: we are now allowed \dneff$=1$, which is in better accord with a ``theory prior" that the sterile neutrino  would be fully populated by oscillations with active
neutrinos for typical mixing angles.  Meanwhile, we see somewhat less residual tension between the early and late Universe, especially
if galaxy clusters are indeed a bit more massive than we have assumed in our main analysis.

 Each of these new ingredients will soon be 
cross checked by a wide variety of upcoming observations. If all are confirmed, observational cosmology will have provided not one but two clear 
discoveries of particle physics beyond the Standard Model within a short space of time, giving long sought clear guidance for how to advance
physics into the future.

\acknowledgements
During completion of this work a similar study appeared \cite{Zhang:2014dxk}; our addition of the
SPT/ACT data sets place stronger upper limits on the allowed \dneff. Ref. \cite{Giusarma:2014zza} also finds preference for an extra relativistic species when adding BICEP2 data, but they do not include the cluster data in their analysis, which constrain the mass.
We thank R. Keisler, M. LoVerde and M. Turner for useful conversations.    
CD was supported by the National Science Foundation grant number AST-0807444, NSF grant number PHY-088855425, and the Raymond and Beverly Sackler Funds. MW was supported by a James Arthur Postdoctoral Fellowship. DHR was supported by the KICP and the Research Computing Center at the University of Chicago.
WH was supported at the KICP through grants NSF PHY- 0114422 and NSF PHY-0551142 and an endowment from the Kavli Foundation, and by U.S. Dept. of Energy contract DE-FG02- 90ER-40560. 

\bibliography{bibbicep}

\end{document}